# Dependence of functional mechanism of matching layer on excitation signal type for ultrasonic transducers

Chunying Wang[a, b, c, *], Cong Wang[a, b, c], Yu Lan[a, b, c] and Wenwu Cao[d]

[a] Acoustic Science and Technology Laboratory, Harbin Engineering University, Harbin 150001, China;

[b] Key Laboratory of Marine Information Acquisition and Security (Harbin Engineering University), Ministry of Industry and Information Technology; Harbin 150001, China;

[c] College of Underwater Acoustic Engineering, Harbin Engineering University, Harbin 150001, China；

[d] Materials Research Institute, The Pennsylvania State University, University Park, Pennsylvania 16802, USA.

## Abstract

Acoustic matching layers are widely employed in high-frequency transducers which are excited by different signal types depending on applications. In this study, a theoretical method has been proposed to investigate the dependence of functional mechanism of matching layer on excitation signal types, i.e., short pulse and long pulse. The results indicate that the matching layer acts as a bandpass frequency filter under the two excitation signal types. In the short pulse excitation case, the matching layer can improve bandwidth and transmitting voltage response simultaneously, whereas for the case of long pulse excitation, the increased bandwidth is at the expense of transmitting voltage response. To verify our theoretical results, underwater acoustic transducers with and without matching layer were fabricated and tested. The thickness design principle of matching layer was modified due to the frequency-dependent acoustic radiation impedance. The experimental results verified the theoretical prediction. Our results provide a deep insight into the fundamental principle of matching layer design according to practical applications.

## 1. Introduction

Matching layer has been used in a wide range of high frequency transducers, including, medical imaging transducers [1, 2], underwater acoustic transducers [3-5], industrial non-destructive evaluation transducers [6-8], etc. These transducers are excited by diverse types of electrical signal depending on their applications. For a medical imaging transducer, a single unipolar pulse is used to achieve a better axial resolution [9]. As for a underwater detection acoustic transducer, tone-burst with multiple cycles are used to trade off the resolution for longer detection distance [10, 11]. It is known that the characteristic impedance of the matching layer is an essential parameter for medical imaging transducer, which is given by $Z_m = \sqrt{Z_w Z_P}$ , where $Z_p$ and $Z_w$ are the characteristic impedance of piezoelectric elements and water, respectively. Whereas, Minoru Toda has proposed a matching layer design concept for narrowband continuous wave devices, where the impedance of matching layer is even lower than that of water [12]. Therefore, it is obvious that the matching layers in the mentioned transducers play their role via different mechanisms. However, most investigations of matching layer design are focusing on medical field and non-destructive evaluation field rather than underwater acoustic field, and the bandwidth and amplitude fluctuation tune method for the transmitting voltage response of underwater transducer is not clear. Moreover, there is no specialized design principle for underwater acoustic transducer related to the acoustic radiation. Up to now, the functional mechanism and design principle for underwater acoustic transducers under different excitation have not been differentiated clearly.

In this work, underwater acoustic transducers with single matching layer were investigated, and the relationship between functional mechanism of matching layer and the excitation signal types was studied in detail. Specifically, the classical KLM (Krimholtz, Leedom and Matthaei) network, which is suitable for designing transducers with multiple matching layers, has been introduced [13]. In the KLM model, emission transfer functions of transducers with or without matching layer were calculated. Subsequently, single pulse signal and long pulse signal were applied to the transducer terminals, then the frequency emission response and impulse response of above-mentioned transducer were calculated. Finally, to verify the theory predictions, prototype transducers composing of self-made 1-3 piezoelectric composites and matching layers were fabricated and tested under long-pulse signal excitation.

## 2. Model building of the transducers with and without matching layer

A typical 1-3 piezoelectric composite transducer is used as an example for the analysis. The 1-3 composite plate with diameter of 30 mm was made of lead zirconate titanate ceramic (PZT-5H) pillar and epoxy (Epotek301, Technology, Inc) kerf filler, the material properties are listed in Table I. The thickness of the plate is 4 mm, and the pillar and kerf width of sample are 0.92 mm and 0.5 mm, respectively. Besides, one quarter-wavelength matching layers was attached to the composite plates, and the characteristic impedance of the matching layer was designed based on either $Z_{\mathrm{mI}} = Z_{\mathrm{p}}^{1/2} Z_{\mathrm{L}}^{1/2}$ [14] or $Z_{\mathrm{mII}} = Z_{\mathrm{p}}^{1/3} Z_{\mathrm{L}}^{2/3}$ [15], where $Z_{\mathrm{p}}$ and $Z_{\mathrm{L}}$ are the characteristic impedances of the piezoelectric plate and water loading, respectively. According to the KLM method shown in Fig. 1(a), the total transfer matrix, which includes one electrical terminal, two mechanical terminals and matching layer, is as follows,

$$T = \begin{bmatrix} \mathrm{T}_{11} & \mathrm{T}_{12} \\ \mathrm{T}_{21} & \mathrm{T}_{22} \end{bmatrix} = T_1 \cdot T_2 \cdot T_3 \cdot T_4 \cdot T_5, \tag{1}$$

where the individual transfer matrix $T_{\mathrm{i}}$ ($i$=1-5) is given in Ref. [14], then the emission transfer function $S_{\mathrm{v}}$ can be calculated from Eq. (2),

$$S_{\mathrm{v}} = \frac{Z_l}{T_{11} Z_l + T_l}, \tag{2}$$

where $Z_{\mathrm{l}}$ is the characteristic impedance of water. The $S_{\mathrm{v}}$ of emission underwater transducers without and with matching layer I or II is represented in Fig. 1(b), together with the applicable bandwidth of these transducers, herein, the applicable bandwidth is defined as frequency range between the half-peak amplitude of the $S_{\mathrm{v}}$ curve has been proposed. As shown in Fig. 1(b), the center frequency of these transducers is all 350 kHz. One can see that the transducer without matching layer possesses the highest peak amplitude and narrowest applicable bandwidth, while the transducer with matching layer I presents a little broader applicable bandwidth, ~ 9%, but shows stronger fluctuation than that of transducer with matching layer II. Therefore, there is a trade-off between applicable bandwidth and amplitude. In the following section, matching layer II will be chosen as an example in our investigation.

Table I. Material parameters of 1-3 PZT5H/epoxy composites

| $c_{33}^{E}$ ($10^{10}$ N/m$^2$) | $e_{33}$ (C/m$^2$) | $\varepsilon_{33}^{S}/\varepsilon_0$ | $\rho$(kg/m$^3$) |
|---|---|---|---|
| 5.19 | 13.4 | 830.6 | 3857 |

The property data of epoxy is calculated according to literature [16].

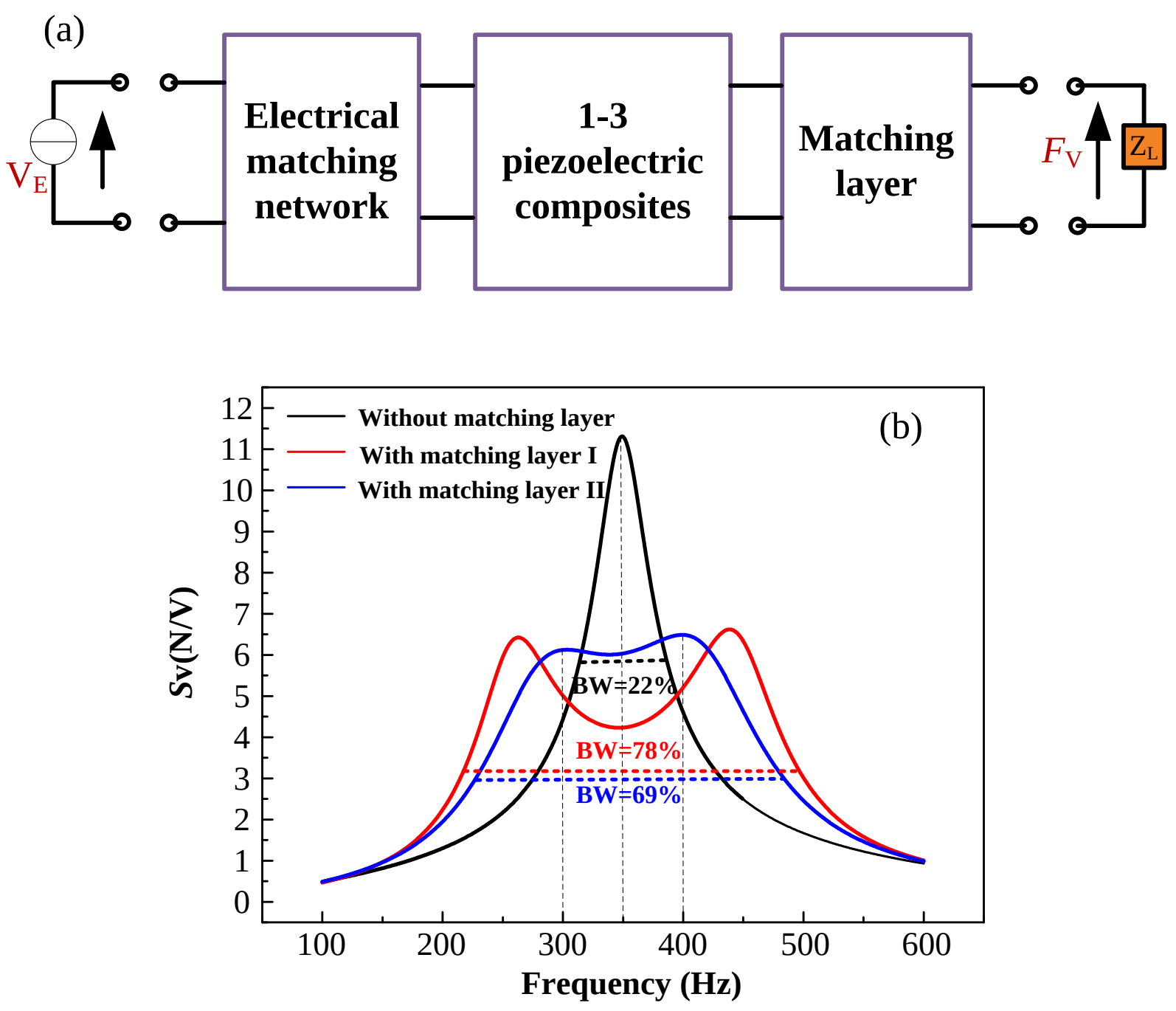

Fig. 1. (a) KLM scheme for the emission broad transducer ($Sv = F_V/V_E$, $V_E$ is the voltage source, and $F_V$ is the force at radiation surface), and (b) the emission transfer function of a typical transducer, without matching layer and with one-quarter wavelength matching layer of different characteristic impedance, matching layer I ($Z_{mI}$=3.17 Mrayl, ρ=1256kg/m$^3$, $v$=2523m/s), and matching layer II ($Z_{mII}$=4.60 Mrayl, ρ=1823kg/m$^3$, $v$=2526m/s), with thickness of 1.9 mm.

Two different excitation signals will be applied on the transducer, i.e., a short- pulse and long pulse consisting of sinusoidal waves at the center frequency. Specifically, the short pulse consists of only 1 cycle, while the long pulse contains 20 cycles, and the amplitude and frequency of the signals are 1V and 350 kHz, respectively. The insert figure in Fig. 2(a) indicates the frequency spectra for both single-pulse and long-pulse signals obtained by means of Fourier transform. Compared with the long pulse signal, the pulse bandwidth of the short pulse is significantly broader. Then the frequency emission response was obtained by the product of the frequency spectra and emission transfer function as shown in Fig. 2. For the short pulse case, the response bandwidth of the transducer with matching layer possesses much broader range than that of its counterpart without the matching layer, hence, the energy spreads over broader frequency range. Meanwhile, the emission impulse response $F_v$ can be obtained by the use of inverse Fourier transform from the frequency emission response. Higher amplitude of the impulse response and less ringdown have been achieved for the transducer with matching layer compared with its counterpart, as shown in Fig. 3(a), therefore, the resolution of transducer with matching layer in the time domain is much better.

For the long pulse case, the emission energy concentrates in a very narrow frequency band. In this case, the maximum amplitude of the frequency emission response is determined by the amplitude of the emission transfer function as shown in Fig. 1(b). For example, at 350kHz, the transducer without matching layer possesses twofold higher amplitude of frequency emission response than that of transducer with matching layer as shown in Fig. 2(b). Therefore, the impulse response of the transducer without matching layer presents a higher amplitude than that of transducer with matching layer as shown in Fig. 3(b), however, it is a different scenario in the cases of 300kHz and 400kHz as shown in Fig. 3(c) and Fig. 3(d)， where the impulse response of the transducer with matching layer presents a higher amplitude than that of transducer without matching layer. Obviously, the conclusions are opposite for the short pulse and long pulse situations between transducers with and without the matching layer.

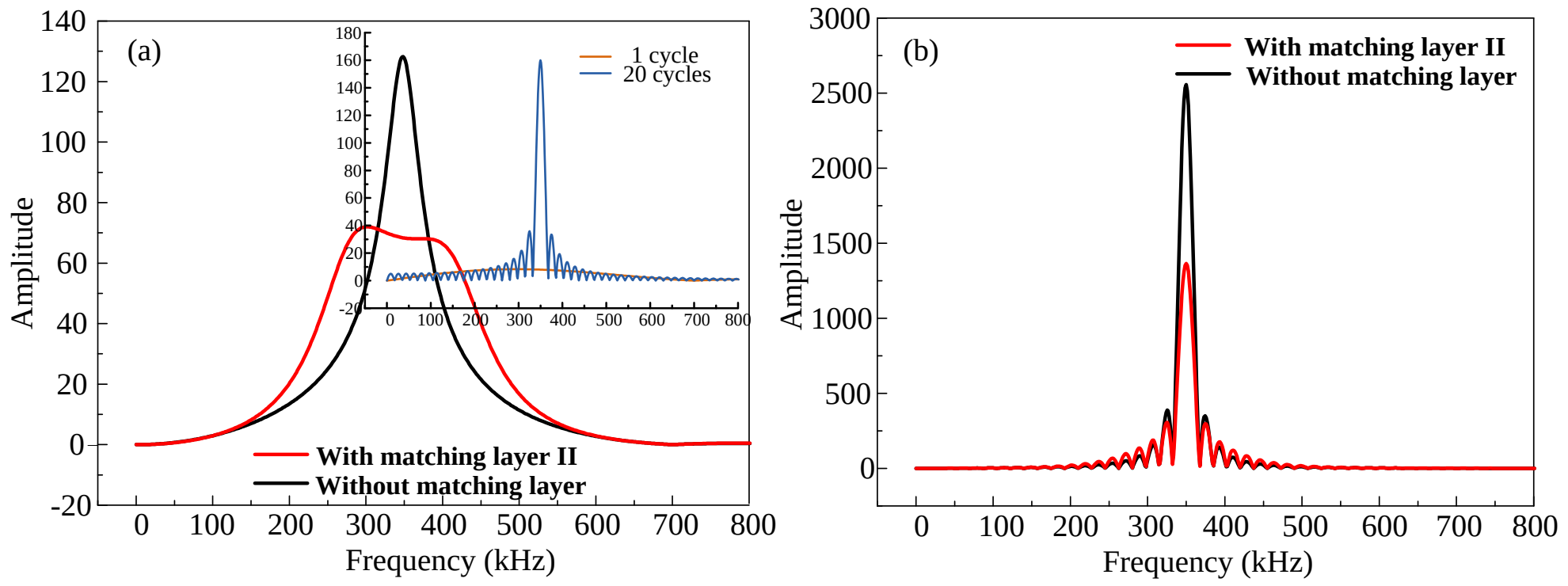

Fig. 2. Frequency emission response of transducers (a)excited by 1 cycle, and (b) 20 cycles (the inset figure represents frequency spectra of short and long pulse signal).

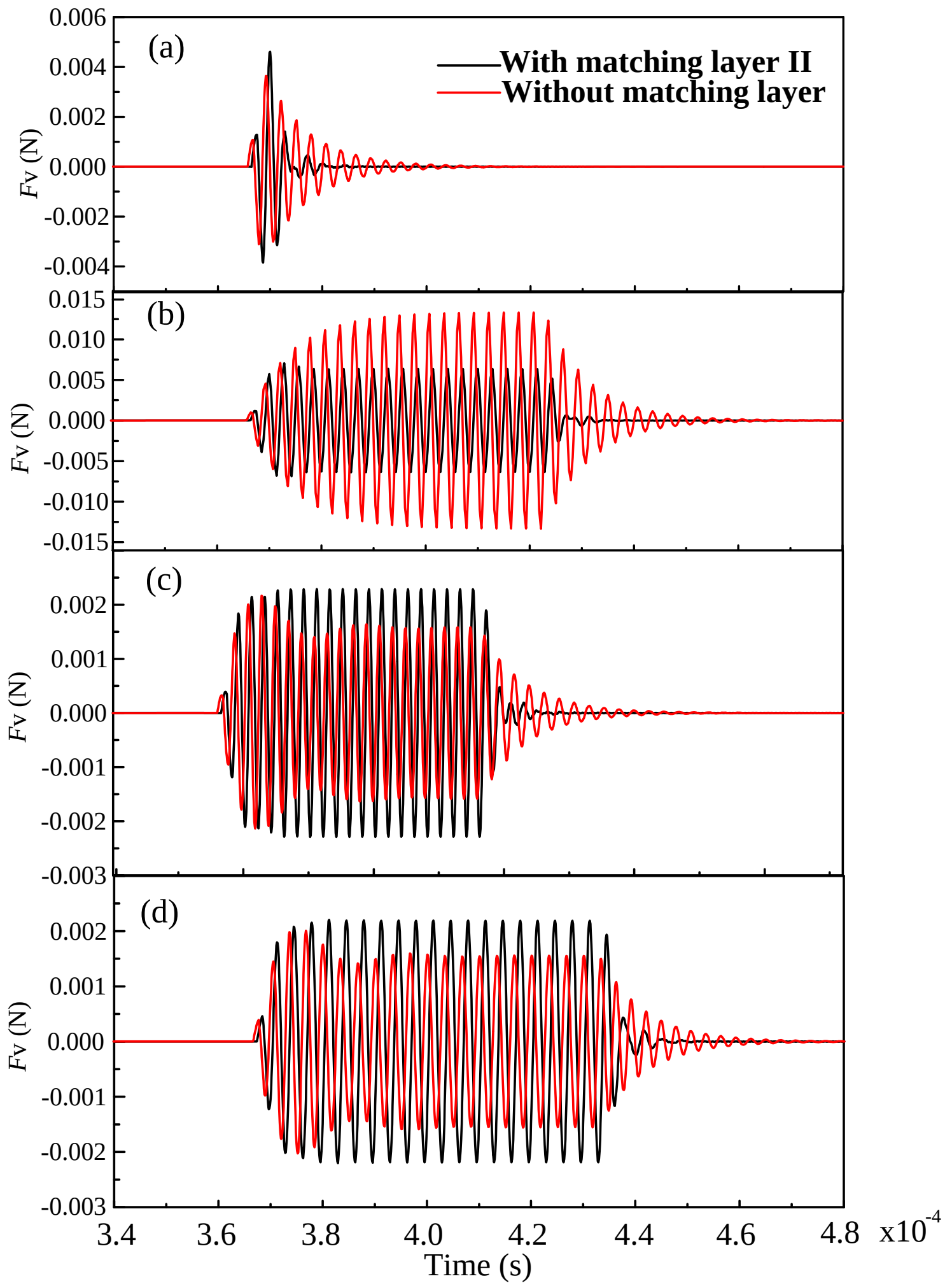

Fig. 3. Impulse emission response of transducers with and without matching layer under (a) 1 cycle pulse excitation and 20 cycle pulses excitation at (b) 350kHz, (c) 300kHz, and (d) 400kHz.

Under short pulse driven situation, the matching layer serves as a bridging medium, and the transducer is in non-resonant state. The functional mechanism is indicated in Fig. 4(a), in this case the acoustic wave reverberates back and forth repeatedly within the matching layer, producing overlap waves which are in phase in the medium, therefore, matching layer can improve the acoustic wave transmission. Besides, the 100% transmission through the matching layer occurs at the frequency corresponding to thickness of one-quarter wavelength, as frequencies removed away the centre frequency, the performance of matching layer will degrade, hence, the matching layer acts as bandpass frequency filter [18]. Whereas, it is noted that the pulse bandwidth cannot be more than that of the transducer itself and there is an upper limit of the bandwidth, and these upper limits are the applicable bandwidth of emission transfer function, i. e. 69% in our case. while for the long pulse situation, the transducer is in resonance state and the matching layer is a component of the whole vibrator. Herein the other vibration mode is introduced due to the additional matching layer, therefore, double peaks correspond to the first and the second order resonances, the schematic of vibration modes is illustrated in Fig. 4(b). In this scenario, only the frequency locating in the applicable range of emission transfer function can excite the transducer effectively, i. e. the fluctuation amplitude range of transmitting voltage response for underwater transdcuer is less than -3dB.

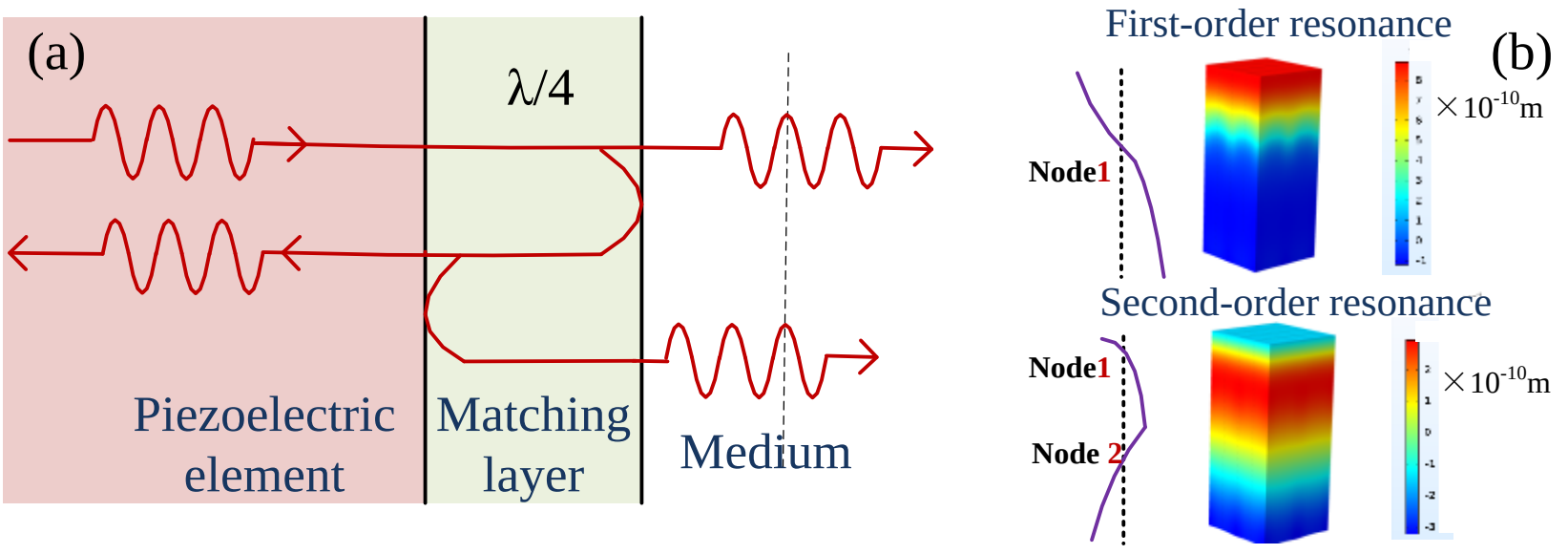

Fig. 3. The functional mechanism of matching layer under different pulse drive situation. (a) short pulse and (b) long pulse (the node figures represent the longitudinal displacement).

## 3. Experimental validation of the high frequency underwater acoustic transducers

To verify our theoretical results, two underwater acoustic transducers with and without matching layer were fabricated. Both transducers consist 1-3 piezoelectric composite plates with the same structure parameters as the theoretical models, the thickness of the former is 3.99 mm ± 0.2 mm and the latter is 3.85mm ± 0.2 mm, respectively. The characteristic impedance of the matching layer was chosen as $Z_{mII}$, and fabricated by epoxy (Epotek 301) resin and15% wt. alumina powder. Obviously, there is a slight disparity between experimental characteristic impedance as showed in Table II and theoretical characteristic impedance $Z_{mII}$ due to the experimental error. The conductance curve in water and the transmission voltage response were measured in a water tank with the dimensions of L 1.5m× W 1.2m× H 1.1m. The two transducers are excited by a 20 cycles sinusoidal-wave signal with the frequency of 350k. We found that 20 cycles excitation is sufficient to approach steady state vibration. Then the receive voltage is obtained by a hydrophone (Reson-4035) and the transmission voltage response is calculated by Eq. (3)

$$\mathrm{TVR_m} = 20\log_{10}(d \cdot V_{out} / V_{in}) - Mel \, , \tag{3}$$

where $V_{out}$ and $V_{in}$ are output voltage amplitude and the receive voltage amplitude by hydrophone, respectively, and $d$=0.5m is acoustic centre distance between the emission transducer and the hydrophone, $Mel$ is the hydrophone sensitivity.

Table II. Material parameters of matching layer

| Velocity(m/s) | Poisson's ration | $\rho$(kg/m$^3$) |
|---|---|---|
| 2644 | 0.35 | 1256 |

These data are measured by ultrasonic pulse-echo method [17].

Due to the fact that acoustic radiation resistance is frequency-dependent and generally increases with frequency, and not a constant value like $Z_l$ in Eq. (2).The acoustic pressure according to Eq. (4) is also frequency-dependent, herein the far field axial acoustic pressure $p(\omega, r)$ of the circular piston transducer is calculated by

$$p(\omega,r)=\frac{j\rho\,\omega\,S}{2\pi r}v(\omega)=\frac{j\rho\,S\,f}{r}v(\omega) \tag{4}$$

where $\omega$ is the angular frequency, $\rho$, $v$ and $S$ and are the density of water, the vibration velocity of radiation surface and radiation area of the transducer, respectively. Besides, $r$ corresponds to the far field distance.

That is to say, the second resonant peak $f_2$ of the conductance curve should be lower than the first resonant peak $f_1$ to compensate the additive acoustic pressure. Therefore, in order to obtain a flat transmitting voltage response curve, the amplitude ratio of the first peak and second peak should be $f_2 / f_1$. To meet the assumption, the thickness of matching layer is reduced to $1.53 \pm 0.1$ mm. Then, the experimental conductance curves of transducers with and without matching layer are normalized in Fig. 4(a). The first and the second resonant peaks of the transducer with matching layer locate at 324kHz and 435kHz, respectively, so that $f_1/f_2$=0.75. Therefore, the amplitude ratio of the conductance peaks should be adjusted to 0.75. However, the practical measured value is 0.69, which is slight smaller than the theoretical value. Then the transmitting voltage response curve can be calculated according to Eq. (5), setting $r$ =1 m.

$$\mathrm{TVR_t} = 20\log\left[\frac{p(r)}{1\mu Pa/V}\right] \tag{5}$$

Both the theoretical and measurement transmitting voltage response curves are shown in Fig. 4(b). The maximum response amplitude of the transducer without matching layer are around 3 dB and 2 dB higher than that of the transducer with matching layer for theoretical and measurement results, which verified the conclusion obtained in the context of Fig. 3(b). It is noted that the second peak amplitude of the TVR is about 1dB lower than that of the first peak due to the smaller conductance amplitude radio 0.69 than theoretical one 0.75. The -3dB theoretical bandwidth for transducers with and without matching layer are 21% and 53%, while the measured ones are 16% and 59%, respectively. Although the exact numerical agreement is hindered by the experimental error during the fabrication of the transducer, the theoretical conclusions are well verified by the experimental results.

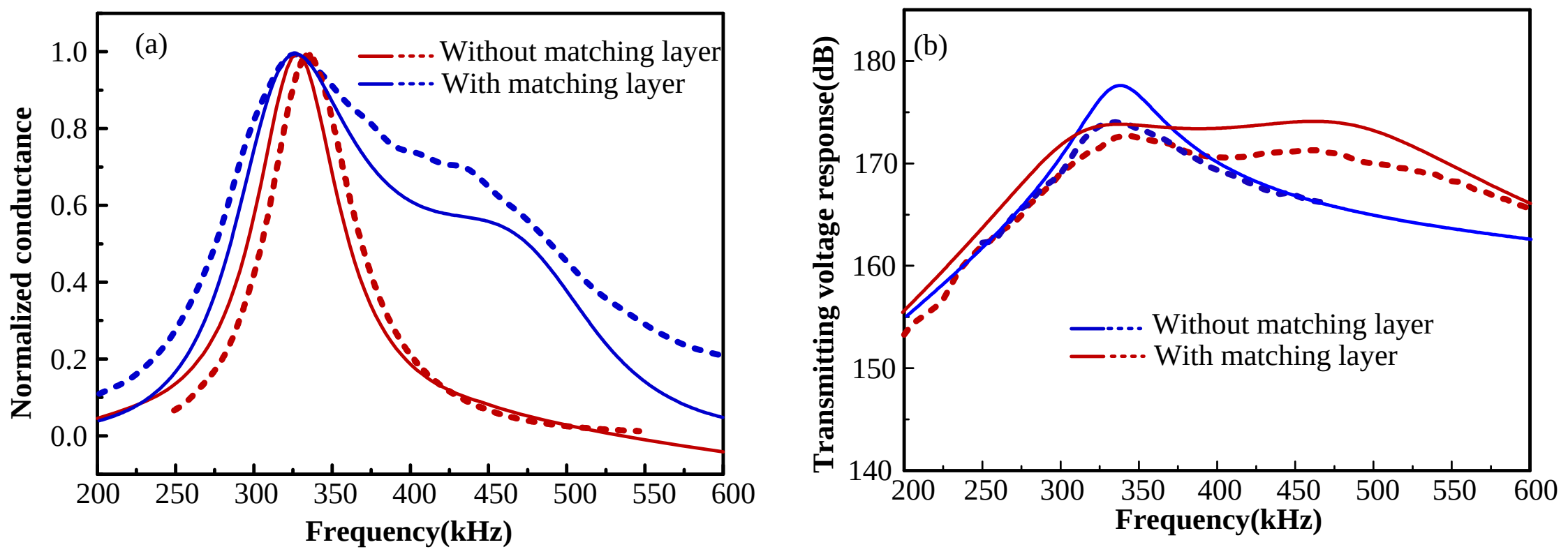

Fig. 4. The theoretical and experimental (a) conductance curve and (b) transmission voltage response for transducers with and without matching layer (the blue solid line, dashed line and the red solid line, dashed line represent the theoretical and measurement result of transducers without and with matching layer).

## 4. Conclusions

In summary, under short pulse excitation, the matching layer serves as a bridging medium and works in non-resonant state, and the introduction of the matching layer can simultaneously improve the bandwidth and transmitting voltage response of the transducer. As for the long-pulse excitation scenario, the matching layer works in non-resonant state and participates the structure vibration, and the increase of the bandwidth is at the expense of the transmitting voltage response. For the both cases, the bandwidth of transducer is determined by the applicable bandwidth of emission transfer function. Then, two underwater acoustic transducers fabricated in our study demonstrated that the characteristic impedance of matching layer determined by $Z_{\mathrm{mII}} = Z_{\mathrm{p}}^{1/3} Z_{\mathrm{L}}^{2/3}$ can give flatter response, and broader bandwidth than that of the transducer without matching layer. Finally, a specific design principle for the matching layer of underwater acoustic transduce was proposed, we show that a flat transmitting voltage response can be obtained when the conductance amplitude ratio of the first resonant peak $f_1$ and second resonant peak $f_2$ equals to the frequency ratio $f_2 / f_1$. Our results also provide a general guiding principle for multiple-matching layers and gradient matching layer designs.

## Acknowledgements

Financial support for this work was provided by the National Key Laboratory Foundation (No. 6142108030507) and the National Nature Science Foundation of China [No.11904066].